# A Sensing Policy Based on Confidence Bounds and a Restless Multi-Armed Bandit Model

Jan Oksanen[†*], Visa Koivunen[†*], H. Vincent Poor[*]
[†]SMARAD CoE, Department of Signal Processing and Acoustics
School of Electrical Engineering, Aalto University, Finland
[*]Department of Electrical Engineering, Princeton University, NJ, USA
Email: jan.oksanen@aalto.fi, visa.koivunen@aalto.fi, poor@princeton.edu

*Abstract*—A sensing policy for the restless multi-armed bandit problem with stationary but unknown reward distributions is proposed. The work is presented in the context of cognitive radios in which the bandit problem arises when deciding which parts of the spectrum to sense and exploit. It is shown that the proposed policy attains asymptotically logarithmic weak regret rate when the rewards are bounded independent and identically distributed or finite state Markovian. Simulation results verifying uniformly logarithmic weak regret are also presented. The proposed policy is a centrally coordinated index policy, in which the index of a frequency band is comprised of a sample mean term and a confidence term. The sample mean term promotes spectrum exploitation whereas the confidence term encourages exploration. The confidence term is designed such that the time interval between consecutive sensing instances of any suboptimal band grows exponentially. This exponential growth between suboptimal sensing time instances leads to logarithmically growing weak regret. Simulation results demonstrate that the proposed policy performs better than other similar methods in the literature.

## I. INTRODUCTION

Cognitive radio (CR) is a promising new technology attempting to give a solution to the scarcity of usable radio spectrum. A cognitive radio network consists of secondary users (SUs) that sense the radio spectrum in hope of finding idle frequencies that they could use for transmission in an agile manner. When the SUs sense that the primary users (PUs) are not using a part of the spectrum the SUs may use those frequencies for data transmission. When the PUs activate again the SUs need to vacate the bands and search for idle spectrum elsewhere.

Depending on the spectral allocation of the PUs different frequency bands may be vacant more often than others. Some bands may have higher bandwidths thus potentially supporting higher data rates. Consequently the CR network would like to focus its sensing on those bands that persistently provide high data rates. Since the expected rates are unknown they need to be learned. This learning problem lends itself to the multi-armed bandit (MAB) problem formulation. Then the objective is to strike a balance between learning the expected data rates while simultaneously exploiting the increasing knowledge about the best band. In MAB problems this is knows as the *exploration-exploitation* trade-off.

In the classical MAB problem a player is faced with $N$ slot machines the $n^{\text{th}}$ one of which produces an unknown expected reward $\mu_n$, $n = 1,...,N$. The player's goal is to select at each time instant a slot machine to be played so that its total payoff will be maximized. The selection of the machine is made according to a policy $\pi$. The analogous counterpart of a slot machine in CR is a frequency band and for the reward it is the achieved data rate.

When the bandits are at rest, the states of those slot machines that are not played do not change. If the rewards are also independent and identically distributed (i.i.d.) in time the optimal policy $\pi^*$ is to play the machine with the highest expected reward $\mu_{n^*}$, ($\mu_{n^*} = \max \mu_n$). In the restless MAB (RMAB) formulation the states of the not-played machines may change, similarly as the state of the spectrum may change regardless of whether it is sensed or not. In the restless case when the rewards are i.i.d. in time the optimal policy is the same as in the rested case. When the rewards are time dependent (e.g. Markovian) the optimal policy is no longer to stay with the single best arm but to select the next arm according to the observed states. However, computing the optimal policy for Markovian rewards in the restless case is in general NP-hard. Hence in the literature a weaker version of the optimal policy has been used. It is known as the best single arm policy [1], [2]. The best single arm policy plays only the arm with the highest stationary expected reward. In this paper the optimal policy always refers to the best single arm policy.

The success of a policy $\pi$ may be measured by its expected regret which is the expected difference between the achieved total payoff obtained with policy $\pi$ and the total payoff achievable by the optimal policy $\pi^*$. In [3] it was shown that for any policy the regret is asymptotically lower bounded by a logarithmic function of the time. Policies achieving asymptotic logarithmic regret rate are called asymptotically efficient.

In this paper an asymptotically efficient sensing policy is proposed. The proposed policy is an index policy consisting of a sample mean term and a confidence term. The sample

J. Oksanen's and V. Koivunen's work was supported by TEKES's (the Finnish Funding Agency for Technology and Innovation) project ENCOR in program TRIAL. H. V. Poor's work was supported by the U.S. National Science Foundation under Grant DMS-1118605.

mean term promotes exploitation whereas the confidence term encourages exploration. The confidence term is not an actual confidence bound in the sense of statistical analysis, but it has the property that the lower the confidence term of a frequency band is, the more likely it is that exploration will take place. It is shown that the proposed policy achieves logarithmic regret asymptotically and our simulation results suggest that logarithmic regret is also achieved uniformly.

The rest of the paper is organized as follows. In Section II we give a brief overview of the related work in the area of multi-armed bandit problems and sensing policies in dynamic spectrum access. In Section III we formulate the problem of finding a spectrum sensing policy mathematically. Sections IV and V present the proposed sensing policy and give analytical results on the asymptotic regret. Section VI illustrates the performance of the policy and verifies the analytical results using simulation examples. The paper is concluded in Section VII.

## II. Related Work

In [3] a class of confidence bound based policies that achieve asymptotic logarithmic regret was presented for the MAB problem. However, these policies are hard to compute and require storing the entire reward history. In [4] a simpler class of asymptotically efficient policies that are easier to compute was derived. In [5] a computationally simple and efficient policy called Upper Confidence Bound (UCB1) was proposed and shown to achieve uniformly logarithmic regret with bounded rewards. Spectrum sensing policies based on the restless multi-armed bandit model have been for example proposed in [1], [2] and [6]. In [6] a myopic index policy was proposed for the RMAB problem for $N$ channels, where the rewards among the bands are independent but obey the same partly known Markov chain. In [1] a policy achieving weak logarithmic regret in the RMAB problem with unknown Markovian rewards was proposed for centralized and decentralized CR networks. In [2] a policy based on deterministic sequencing of exploration and exploitation epochs was proposed for the RMAB problem and shown to achieve logarithmic weak regret.

The policy proposed in this paper resembles the one in [2] in the sense that the periods of exploitation in the proposed policy also tend to grow exponentially. However, the policy in this paper attains a simpler form and according to our simulation also yields better performance when the rewards are bounded. The novelty of the proposed policy is in the confidence term that is designed so that the expected time interval between two consecutive sensings of a suboptimal band grows exponentially. This exponential growth in time is shown to result in asymptotically logarithmic weak regret to be defined below.

## III. Problem Formulation

### A. System model

At time instant $t$ a CR network senses (and possibly accesses) frequency band $n, n = 1, ..., N$, and achieves a random reward (data rate) $x_n(t)$ with an unknown mean $\mu_n$. We may think that $x_n(t) = (1 - S_n(t))r_n(S_{n(t)})$, where $S_n(t) \in [0, 1]$ is the random state of band $n$ at time $t$ and $r_n(S_n(t))$ is the data rate obtained from band $n$ when the band is in state $S_n(t)$. In this paper $S_n(t) = 1$ means that the band is occupied and $S_n(t) = 0$ that it is idle. The state of the band is assumed to evolve according to a stationary 2-state Markov chain or according to an i.i.d. Bernoulli process. The observed rewards $x_n(t)$ are assumed to be bounded within $[0, 1]$ by scaling the observed data rates by the inverse of the highest Shannon capacity among the bands. It is assumed that the CR network has a way to estimate and give feedback about the achieved rates to a central node, e.g. a fusion center. The rewards among the bands are assumed to be independent.

### B. Objective

The CR's objective is to find a sensing policy that achieves an optimal exploration-exploitation trade-off. The success of a policy $\pi$ is measured by its expected regret $\mathrm{E}\big[R^\pi(t)\big]$. The regret of a policy $\pi$ is defined as the difference of the total payoff achieved by the policy $\pi$ and the total payoff achievable by the optimal policy $\pi^*$ as

$$\mathrm{E}\big[R^\pi(t)\big] = t\mu_{n^*} - \sum_{n=1}^{N} \mu_n \mathrm{E}\big[T_n^\pi(t)\big], \qquad (1)$$

where $t$ is the time index, $T_n^\pi(t)$ is the number of times band $n$ has been sensed up to time $t$ using policy $\pi$ and $n^* = \arg\max_n \mu_n$. In order to simplify the notation the superscript $\pi$ will be dropped for the rest of the paper. In case the rewards are dependent over time, e.g. when they obey a Markov model, the optimal policy $\pi^*$ is termed the best single band policy and the definition in (1) is then known as the *weak regret*. In this paper only weak regret is considered.

### C. Practicality the System Model

In practice the notion of best single subband is ambiguous. Since the SUs may be scattered in space they experience different channel fading and consequently obtain different data rates in different locations at a given frequency band. For the same reasons the probabilities of detection and false alarm in spectrum sensing may vary across the bands. Taking these issues into account the optimal sensing policy becomes a function of the access policy (i.e., who will get access to the possibly idle band) and the employed sensing scheme. Such joint optimization of sensing and access in CR has been considered for example in [7], [8] and [9].

In most imaginable CR scenarios the rewards (the data rates) are not stationary. Among other factors the obtained rates depend on the traffic load in the primary network that may vary between peak and off-peak hours. In such cases policies such as the $\epsilon$-greedy [8] or discounted UCB [10] should be considered.

## IV. The Proposed Policy

In this section we propose a spectrum sensing policy for CR with stationary but unknown reward distributions. The

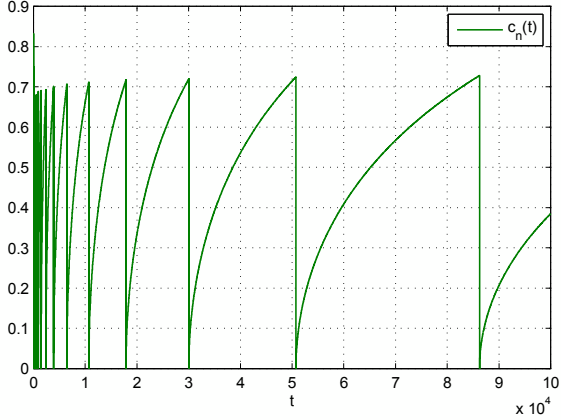

Fig. 1. The confidence term of a suboptimal channel as a function of time. The time interval between two consecutive sensing instances (the zeros of $c_n(t)$) of the suboptimal band tend to grow exponentially in time. Since the time intervals between two consecutive sensings of a suboptimal channel grows exponentially it means that the total number of sensings must grow logarithmically in time.

proposed sensing policy is as follows:

1. Sense each subband once.
2. For $t > N$ sense the band with the highest index $I_n(t)$ where
$$I_n(t) = \bar{x}_n(t) + c_n(t). \quad (2)$$

In (2) $\bar{x}_n(t) = \frac{1}{T_n(t)} \sum_{m=1}^{t} x_n(m)$ is the reward sample mean of band $n$. The confidence term is given by

$$c_n(t) = \sqrt{\ln\left(\frac{t}{\tau_n(t)}\right)}, \quad (3)$$

where $\tau_n(t)$ is the last time instant when band $n$ was sensed.

It should be noted that $c_n(t)$ is not an actual confidence bound in the sense of statistical analysis. This can be easily noticed from the facts that $c_n(t)$ does not depend on the sample size and that $c_n(t) = 0$, when $\tau_n(t) = t$. The term $c_n(t)$ does however reflect confidence in the sense that the longer time ago the band was sensed the larger the $c_n(t)$ will be. Figure 1 shows the behavior of the confidence term of a suboptimal subband as a function of time. It can be observed that the confidence term does not converge as more samples are gathered as it would if it were an actual confidence bound. However, it has the property that the higher the confidence term is the more likely it is that exploration will take place. Furthermore, it can be seen that the zeros of the confidence term mark the sensing instances of the subband and that the time intervals between these instances tend to grow exponentially. This observation is the basis of the regret analysis in the next section.

## V. REGRET ANALYSIS

In this section we show that the proposed policy attains asymptotically logarithmic regret. As was seen in (1) the expected regret depends on the number of times that a suboptimal frequency band is sensed. In order to show that the regret is logarithmic one needs to show that the expected number of sensings on any suboptimal band is upper bounded logarithmically. Our analysis is based on inspecting the finite interval between two consecutive sensing instances of any suboptimal band (the instances of the zeros seen in Figure 1). It will be shown that the time difference between the $k^{\text{th}}$ sensing instance of a suboptimal band and the first sensing instance grows exponentially as a function of $k$ which consequently leads to logarithmic regret.

Assume that a suboptimal band, $n \neq n^*$, has been sensed for the $k^{\text{th}}$ time at time instance $z_{n,k}$. The next sensing instance of the band is a random variable $z_{n,k+1}$. By showing that given the previous sensing instance $z_{n,k}$ there exists a constant $C_n > 1$ such that

$$\mathrm{E}\left[z_{n,k+1} \Big| z_{n,k}\right] \geq z_{n,k} C_n, \quad (4)$$

one can conclude that the policy has logarithmic expected regret. Conditioning with respect to $z_{n,k}$ is needed since as can be seen in Figure 1 the rate of the increase in $c_n(t)$ depends on the previous sensing time. Logarithmic regret can be shown to follow from (4) by assuming that (4) is true and using the law of iterated expectations. Denote the time instance of the first sensing of the suboptimal band $n$ as $z_{n,1}$. Assuming that (4) is true the expected time instance of the second sensing given $z_{n,1}$ will be $E[z_{n,2}|z_{n,1}] \geq z_{n,1} C_n$. Similarly for the third sensing instance $\mathrm{E}[z_{n,3}|z_{n,1}] \geq z_{n,1} C_n^2$ and so on for the $k^{\text{th}}$ sensing instance $t \doteq \mathrm{E}[z_{n,k}|z_{n,1}] \geq z_{n,1} C_n^{(k-1)}$. Taking the logarithm of both sides yields that the number of sensings of a suboptimal band $n$ at time $t$ is

$$k \leq \frac{\log(t) - log(z_{n,1})}{\log(C_n)} + 1, \quad (5)$$

where $C_n > 1$ is a constant. Next we show that (4) holds asymptotically for the proposed policy. Intuitive analysis is given in Section V-A and a more detailed analysis for i.i.d. rewards and Markovian rewards is given in Sections V-B and V-C respectively.

### A. Intuitive Sketch of Proof of Asymptotic Logarithmic Regret

In this section we give an intuitive sketch of a proof that the proposed policy attains asymptotic logarithmic regret. Assume that suboptimal band $n$ was sensed for the $k^{\text{th}}$ time at time instant $z_{n,k}$. According to the law of large numbers as more and more samples are obtained from the optimal and the suboptimal band the better the indices may be approximated as

$$\begin{aligned} I_n(t) &\approx \mu_n + c_n(t) \\ I_{n^*}(t) &\approx \mu_{n^*} + c_{n^*}(t) \geq \mu_{n^*}. \end{aligned}$$

The later inequality follows from the fact that the confidence term $c_{n^*}(t) \geq 0$. The suboptimal band $n$ will not be sensed

sooner than when $I_n(t) \geq \mu_{n^*}$. This happens when

$$\mu_n + \sqrt{\ln(t/z_{n,k})} \geq \mu_{n^*}$$
$$\Rightarrow z_{n,k+1} \geq t \geq z_{n,k} e^{\Delta_n^2}, \quad (6)$$

where $\Delta_n = \mu_{n^*} - \mu_n$ is the optimality gap of band $n$ and $z_{n,k+1}$ is the next sensing instance. Now it can be seen that $e^{\Delta_n^2} > 1$ is a constant and that (6) is of the form of the requirement in (4) given for logarithmic regret. Since the expected next sensing instance of a suboptimal band $n$ is always greater than or equal to its previous sensing instance times a constant ($e^{\Delta_n^2} > 1$), the time interval between consecutive sensing grows exponentially as a function of the number of sensings. Hence, the total number of sensing instances of a suboptimal band is asymptotically bounded by a logarithmic function.

*B. Independent Rewards*

Assume $N$ frequency bands each with i.i.d. rewards, in which the rewards in the $n^{\text{th}}$ band have expected value $\mu_n$, $n = 1, ..., N$. Denote the optimality gap as $\Delta_n = \mu_{n^*} - \mu_n$. Given that a suboptimal band $n$ has been sensed for the $k^{\text{th}}$ time at time instance $z_{n,k}$ the expected time instance when it will be sensed again is $\mathrm{E}[z_{n,k+1}|z_{n,k}]$. Note that $z_{n,k}$ is also a discrete time index and that it is of the same scale as $t$. The conditional expectation may be partitioned into a sum of two mutually exclusive events:

$$\begin{aligned}\mathrm{E}\big[z_{n,k+1}\big|z_{n,k}\big] &\geq \mathrm{E}\big[z_{n,k+1}\big|z_{n,k}, \bar{x}_n(z_{n,k}) \geq \bar{x}_{n^*}(z_{n,k})\big] \\ &\quad \cdot \mathrm{P}\big\{\bar{x}_n(z_{n,k}) \geq \bar{x}_{n^*}(z_{n,k})\big\} \\ &\quad + \mathrm{E}\big[z_{n,k+1}\big|z_{n,k}, \bar{x}_n(z_{n,k}) < \bar{x}_{n^*}(z_{n,k})\big] \\ &\quad \cdot \mathrm{P}\big\{\bar{x}_n(z_{n,k}) < \bar{x}_{n^*}(z_{n,k})\big\}. \quad (7)\end{aligned}$$

The inequality in (7) comes from the fact that for band $n$ to be selected for sensing its index must be at least as big as the index to the optimal band $n^*$. When $\bar{x}_n(z_{n,k}) \geq \bar{x}_{n^*}(z_{n,k})$ the expected time instance when band $n$ is sensed again is

$$\mathrm{E}\big[z_{n,k+1}\big|z_{n,k}, \bar{x}_n(z_{n,k}) \geq \bar{x}_{n^*}(z_{n,k})\big] \geq z_{n,k} + 1. \quad (8)$$

When $\bar{x}_n(z_{n,k}) < \bar{x}_{n^*}(z_{n,k})$ we may use Jensen's inequality for convex functions to bound $\mathrm{E}\big[z_{n,k+1}\big|z_{n,k}, \bar{x}_n(z_{n,k}) < \bar{x}_{n^*}(z_{n,k})\big]$. The next sensing of band $n$ occurs when $I_n(t) \geq I_{n^*}(t)$, which happens no sooner than the event

$$\bar{x}_n(t) + \sqrt{\ln(t/z_{n,k})} \geq \bar{x}_{n^*}(t).$$

Solving for $t$ and taking the expectation from both sides yields

$$\begin{aligned}&\mathrm{E}\big[z_{n,k+1}\big|z_{n,k}, \bar{x}_n(z_{n,k}) < \bar{x}_{n^*}(z_{n,k})\big] \\ &\geq \mathrm{E}\big[t \big| z_{n,k}, \bar{x}_n(z_{n,k}) < \bar{x}_{n^*}(z_{n,k})\big] \geq z_{n,k} e^{\Delta_n^2}, \quad (9)\end{aligned}$$

where the latter inequality follows from Jensen's inequality for convex functions. Combining (8) and (9) into (7) we obtain

$$\begin{aligned}\mathrm{E}\big[z_{n,k+1}\big|z_{n,k}\big] &\geq (z_{n,k}+1)\mathrm{P}\big\{\bar{x}_n(z_{n,k}) \geq \bar{x}_{n^*}(z_{n,k})\big\} \quad (10) \\ &\quad + z_{n,k} e^{\Delta^2}\Big(1 - \mathrm{P}\big\{\bar{x}_n(z_{n,k}) \geq \bar{x}_{n^*}(z_{n,k})\big\}\Big).\end{aligned}$$

Now we show that $\mathrm{P}\{\bar{x}_n(z_{n,k}) \geq \bar{x}_{n^*}(z_{n,k})\} \stackrel{t\to\infty}{\to} 0$ and consequently $\mathrm{E}[z_{n,k+1}|z_{n,k}] \stackrel{t\to\infty}{\to} z_{n,k} e^{\Delta^2}$. For i.i.d. rewards this can be shown using the following lemma due to Hoeffding [11]:

**Lemma 1.** *If $Y_1, ..., Y_m$ and $Z_1, ..., Z_n$ are independent random variables with values in the interval $[0,1]$, and if $\bar{Y} = (Y_1 + \cdots + Y_m)/m$ and $\bar{Z} = (Z_1 + \cdots + Z_n)/n$, then for $\epsilon > 0$*

$$\mathrm{P}\big\{\bar{Y} - \bar{Z} - \mathrm{E}[Y] + \mathrm{E}[Z] \geq \epsilon\big\} \leq \exp\Big(-\frac{2\epsilon^2}{m^{-1}+n^{-1}}\Big). \quad (11)$$

*Proof:* (11) follows from Theorem 2 in [11]. ∎

The intuition given by Lemma 1 related to our analysis is that the probability of the sample mean of a suboptimal band being greater than or equal to the sample mean of the optimal band approaches 0 as more and more samples are obtained from both bands. Consequently the first term of the sum on the right-hand-side of (10) approaches 0 as a function of time. Using Lemma 1 one can now bound the probability

$$\begin{aligned}&\mathrm{P}\big\{\bar{x}_n(z_{n,k}) \geq \bar{x}_{n^*}(z_{n,k})\big\} \\ &= \mathrm{P}\big\{\bar{x}_n(z_{n,k}) - \bar{x}_{n^*}(z_{n,k}) - (\mu_n - \mu_{n^*}) \geq \Delta_n\big\} \\ &\leq \exp\bigg(\frac{-2\Delta_n^2}{T_n(z_{n,k})^{-1} + T_{n^*}(z_{n,k})^{-1}}\bigg),\end{aligned}$$

where the inequality follows from Lemma 1. Since the rewards are assumed to be bounded within $[0,1]$ and $c_n(t)$ is an increasing function between two consecutive sensing events (see Figure 1), it can be easily understood that each band will be sensed infinitely many times as $t \to \infty$. Hence, $T_n(t) \stackrel{t\to\infty}{\to} \infty$ and $T_{n^*}(t) \stackrel{t\to\infty}{\to} \infty$, and consequently $\mathrm{P}\{\bar{x}_n(z_{n,k}) > \bar{x}_{n^*}(z_{n,k})\} \stackrel{t\to\infty}{\to} 0$. Then from (10) it follows that $\mathrm{E}[z_{n,k+1}|z_{n,k}] \stackrel{t\to\infty}{\to} z_{n,k} e^{\Delta^2}$. According to (5) the proposed policy attains then an asymptotic expected regret that is $O(\log t)$.

*C. Markovian Rewards*

We can show that the policy attains asymptotically logarithmic weak regret also when the rewards are Markovian. Under some mild conditions on the Markov chain this may be done using the following concentration inequality by Lezaud [12]:

**Lemma 2.** *Let $Y_1, ..., Y_m$ be an aperiodic irreducible finite-state Markov chain with transition matrix $P$. Further, let $\lambda_2$ be the second largest eigenvalue of the multiplicative symmetrization of $P$, and $\mu_Y$ be the expectation of $Y_i$ under the stationary distribution. Then for any $\epsilon \in [0,1]$ there exists*

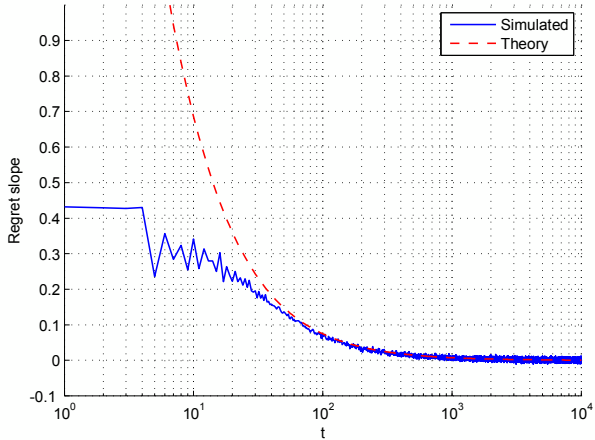
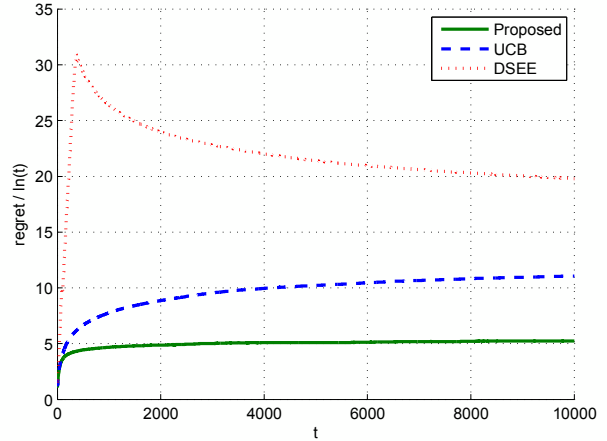

Fig. 2. The simulated slope of the regret curve and the theoretical slope given by (13) for a case of 5 subbands with Markovian rewards. The rewards are $r_n(0) = 1$ and $r_n(1) = 0.1$ for all $n$ and the state transition probabilities are $P_{10} = [0.1, 0.1, 0.5, 0.1, 0.1]$ and $P_{01} = [0.2, 0.3, 0.1, 0.4, 0.5]$. The initial states of the bands are drawn from the stationary distributions. It can be seen that the true regret growth rate approaches the growth rate of the logarithmic curve.

Fig. 3. Regret of the proposed policy, UCB1 and DSEE with Markovian reward. The regret curves are normalized by a factor $1/\ln(t)$. $N = 5$ and the transition probabilities are $P_{10} = [0.1, 0.1, 0.5, 0.1, 0.1]$ and $P_{01} = [0.2, 0.3, 0.1, 0.4, 0.5]$. The initial states of the bands are drawn from the stationary distributions..The rewards are $r_n(0) = 1$ and $r_n(1) = 0.1$ for all five bands. It can be seen that the proposed policy outperforms the two other policies.

a constant $L > 0$ such that

$$P\left\{|\bar{Y} - \mu_Y| \geq \epsilon\right\} \leq L \exp\left(\frac{-m\epsilon^2(1 - \lambda_2)}{28}\right). \quad (12)$$

*Proof:* See [12] page 857. ∎

Similarly to the bounded i.i.d. case, in the finite-state Markovian case the number of sensings $T_n(t)$ tends to infinity as $t \to \infty$. Then according to Lemma 2 $P\{\bar{x}_n(z_{n,k}) > \bar{x}_{n^*}(z_{n,k})\} \stackrel{t\to\infty}{\to} 0$. Consequently from (10) and (5) it then follows that with Markovian rewards the policy will also have asymptotic weak regret that is $O(\log t)$.

## VI. SIMULATIONS

### A. Regret Growth Rate

To verify our analysis in Section V we compare the theoretical logarithmic regret growth rate with the simulated growth rate. Using the result derived in Section V we can calculate the theoretical asymptotic slope of the regret. Taking the derivative of (5) with $C_n = e^{\Delta_n^2}$ with respect to $t$ and summing over all subbands, the asymptotic slope of the regret is given by

$$\frac{d}{dt}E[R(t)] \stackrel{t\to\infty}{\to} \sum_{\substack{n=1 \\ n \neq n^*}}^{N} \frac{1}{t\Delta_n}. \quad (13)$$

Figure 2 plots the derivative of the simulated regret curve and the derivative of the theoretical regret given by (13) when there are 5 bands producing Markovian rewards. In this case it can be seen that the slopes of the true regret curve and the theoretical slope match well in expectation when $t > 100$.

### B. Expected Regret

Figure 3 shows the regret of the proposed policy with Markovian rewards when the number of subbands is $N = 5$. The transition probabilities of the Markov chains are $P_{10} = [0.1, 0.1, 0.5, 0.1, 0.1]$ and $P_{01} = [0.2, 0.3, 0.1, 0.4, 0.5]$ and the rewards are $r_n(0) = 1$ and $r_n(1) = 0.1$. The curves are averages of 10000 independent runs. The regret curves are also shown for two other policies, namely the UCB1 [5] and Deterministic Sequencing of Exploration and Exploitation (DSEE) [2] (with parameter $D = 10$). Both these policies have been shown to achieve uniform logarithmic regret. In the case of Figure 3 the proposed policy outperforms the UCB1 and DSEE policies. Since both the UCB1 and the DSEE have been shown to be uniformly bounded from above by a logarithmic function, the figure shows that in this scenario the proposed policy must also attain uniformly logarithmic regret. The regret curves with i.i.d. Bernoulli rewards with the same stationary mean as in Figure 3 are in this case practically the same as with Markovian rewards, and hence are not shown here.

Figure 4 shows the expected regrets of the three policies for i.i.d. Bernoulli rewards when the probabilities of the bands being idle are $[0.3, 0.36, 0.17, 0.25, 0.33]$. The parameter $D$ in the DSEE is set to $D = \ln(t)$. In this case the UCB1 policy seems to perform the worst, whereas the DSEE and the proposed policy perform equally well.

## VII. CONCLUSIONS

In this paper a sensing policy achieving asymptotic logarithmic weak regret has been proposed for the restless multi-armed bandit problem that arises in spectrum sensing and access in CR. The proposed policy is an index policy consisting of a sample mean term and a confidence term. The confidence term has been chosen such that the average time interval between two sensings of any suboptimal band grows exponentially in time. We have shown using analytical tools that the proposed policy achieves logarithmic weak regret asymptotically. Log-arithmic weak regret in case of i.i.d. and Markovian rewards has been verified in the simulations. Furthermore, simulation results show that the proposed policy achieves in most cases lower regret than the DSEE and UCB1 policies.

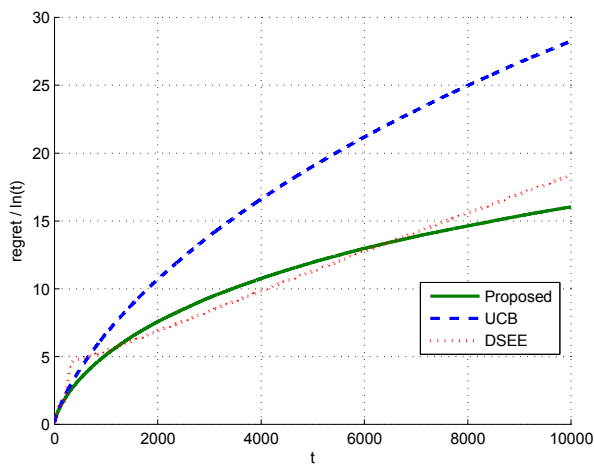

Fig. 4. Regret of the proposed policy, UCB1 and DSEE with i.i.d. Bernoulli rewards. The regret curves are normalized by a factor $1/\ln(t)$. $N = 5$ and the probabilities of the bands being idle are $P_0 = [0.3, 0.36, 0.17, 0.25, 0.33]$. The rewards are $r_n(0) = 1$ and $r_n(1) = 0.1$ for all $n$. In this scenario the regret of the UCB1 policy is clearly higher than that of the DSEE and the proposed policy.